\documentclass{sbrt2017eng}

\usepackage[brazil]{babel}  
\usepackage{enumitem}
\usepackage{mathrsfs,amsmath}
\usepackage{amssymb,amsfonts}
\usepackage{hyperref}
\usepackage{graphicx}
\newcommand{\doi}{doi:}
\renewcommand*{\doi}[1]{\href{http://dx.doi.org/#1}{doi: #1}}

\begin{document}
\title{The Hamming and Golay \\ Number-Theoretic Transforms}

\author{A. J. A. Paschoal \thanks{A. J. A. Paschoal, Pernambuco Federal Institute of Education, Science and Technology (IFPE), Caruaru-PE, Brazil. E-mail: arquimedes.paschoal@caruaru.ifpe.edu.br.}, R. M. Campello de Souza \thanks{R. M. Campello de Souza, Federal University of Pernambuco (UFPE), Recife - PE, Brazil. E-mail: ricardo@ufpe.br.} and H. M. de Oliveira \thanks{H. M. de Oliveira, Statistics department (UFPE), Recife-PE, Brazil. E-mail: hmo@de.ufpe.br.}}

\maketitle

\begin{abstract}
New number-theoretic transforms are derived from known linear block codes over finite fields. In particular, two new such transforms are built from perfect codes, namely the \textit {Hamming number-theoretic transform} and the \textit {Golay number-theoretic transform}. A few properties of these new transforms are presented.
\end{abstract}

\begin{keywords}
Number-theoretic transforms (NTT), Finite fields, Linear block codes, Perfect codes, Perfect  transforms.
\end{keywords}
\section{Preliminaries}

Digital transforms are an interesting  way to construct linear block codes for error correction \cite{Campello de Souza 2009}, \cite{Campello de Souza 2011}. Fourier and Hartley codes were defined associated with the Fourier number-theoretic transform and the Hartley number-theoretic transform, respectively. Recently, a new family of multilevel codes -- the Pascal codes \cite{Paschoal 2018} - has been proposed, which are constructed from the Pascal number-theoretic transform \cite{Paschoal 2015}. Also recently, Paschoal \cite{Paschoal 2018} proposed a reverse way to conceive new digital transforms from known linear block codes. Here we present the technique used to engender two new transforms over finite fields, namely the Hamming Number-Theoretic Transform (HamNT) and the Golay Number-Theoretic Transform (GolNT). From now on these transforms, based on Perfect Codes, are referred to as ``Perfect Transforms''. Tiet\"av\"ainen demonstrated in 1971 that the only existing perfect block codes over $GF(p)$ are \cite{Tietavainen 1973} :\\
\begin{description}
\item[$\bullet$] Hamming codes $\left ( N=\frac{p^m-1}{p-1},~k=\frac{p^m-1}{p-1}-m, ~~d=3 \right )$.\\
\item[$\bullet$] Binary Golay code $(N=23,~k=12,~~d=7)$.\\
\item[$\bullet$] Ternary Golay code $(N=11,~k=6,~~d=5)$.
\end{description}
The perfect codes are also deeply linked to the lattices in Euclidean spaces \cite{Conway-Sloane 2013}. Only the $E_8$ Gosset lattice (associated with the extended Hamming code) and the Leech lattice $\Lambda_{24}$ (associated with the extended binary Golay code) are perfect sphere packings.
\section{New finite field transforms derived from block codes}

In \cite{Campello de Souza 2009} Campello de Souza and coworkers used the fact that for every linear transform $T$, its \textit {eigenvectors} $v$ satisfy
\begin{equation}
\left[T-\lambda \mathbb{I} \right]v = 0,
\end{equation}
\noindent where $\lambda$ denotes the eigenvalue associated with $v$, to show that it is possible to get an associated parity-check matrix for a block code simply by row reducing the  matrix $\left[T-\lambda \mathbb{I} \right]$. In this paper we go on a somewhat opposite direction, i.e., starting from a linear block code, it is possible to derive a new linear transformation (associated with the generating block code) by completing the parity-check matrix, adding $k$ rows, so as to form a square matrix $H_e$ and adding it to the matrix $\lambda \mathbb{I}$, i.e., 

\begin{equation}
T=H_e+\lambda \mathbb{I}.
\end{equation}

Indeed, $\lambda$ must be chosen so as to guarantee that $\det T \neq 0$. Such transform receives the name of the linear block code whose parity matrix was used to derive it.\\
The number of vectors on the whole space of all the $N$-tuples over $GF(p)$ is $p^N$ (with no redundancy). The set of all eigenvectors of a transform of length $N$ defined over a finite field always engender a subspace denoted by $\mathbb{V}$, with dimension $k < N$ with $GF(p)$-valued components.

\begin{definition}
\label{def1}
(perfect transform). A transform $T$ of length $N$ over $GF(p)$ is said to be perfect \textit{if and only if} there is an integer $t$ such that the eigenvectors subspace $\mathbb{V}$ of $T$ has dimension meeting the sphere-packing bound
\begin{equation}
\dim \mathbb{V}=k=N-\log_p\left [ \sum_{i=0}^{t} (p-1)^i\binom{N}{i}\right ].
\end{equation}
\end{definition}\vspace{6pt}

The eigenvector subspace of $T$ is a perfect block code \cite{Tietavainen 1973} when $T$ is a perfect transform.

\section{The Hamming Number-Theoretic Transform}

In this section, we illustrate the aforementioned technique splitting it in two ways:

\begin{enumerate}[label=\alph*)]
    \item Fulfilling the Hamming parity-check matrix by adding null rows or  using linear combinations of its own rows.
    \item Cyclic rotating the parity polynomial, $h(x)$.
\end{enumerate}

\subsection{The Standard Hamming Number-Theoretic Transform}

Consider the binary Hamming code $\mathcal{H}(7, 4, 3)$, with parity-check matrix (reduced row echelon form)

\begin{displaymath}
\arraycolsep=2pt
\renewcommand{\arraystretch}{0.6}
H=
\left[ \begin{array}{lllllll}
1&1&0&1&1&0&0\\
1&1&1&0&0&1&0\\
1&0&1&1&0&0&1\\
\end{array}
\right].
\end{displaymath}

In this case, in order to construct the Hamming Number-Theoretic Transform (HamNT) matrix, $T_{HamNT}^{(\lambda)}$, we need to add four rows to this parity matrix to get the matrix $H_e$. Such additional rows are merely null rows. This new matrix, named $H_e$, is added to $\lambda \mathbb{I}_N$, where $\lambda =1$ is the eigenvalue over $GF(2)$ and $\mathbb{I}_N = \mathbb{I}_7$ is the identity matrix of dimension 7. Note that, in fact, the eigenvalues are unknown and we should test all the possible candidates eliminating all those that result in a singular transformation matrix. The different eigenvalues may produce different transformation matrices. From the binary Hamming code, we get the $7 \times 7$ binary HamNT whose transformation matrix is
\\
\begin{displaymath}
\renewcommand{\arraystretch}{0.6}
T_{HamNT}^{(1)} = 
\left[ \begin{array}{lllllll}
0&1&0&1&1&0&0\\
1&0&1&0&0&1&0\\
1&0&0&1&0&0&1\\
0&0&0&1&0&0&0\\
0&0&0&0&1&0&0\\
0&0&0&0&0&1&0\\
0&0&0&0&0&0&1\\
\end{array}
\right].
\end{displaymath}
The eigenvector matrix is given by
\begin{displaymath}
\renewcommand{\arraystretch}{0.6}
\text{eig } T_{HAM}^{(1)} =
\left[ \begin{array}{lllllll}
1&1&0&0&0&0&1\\
1&1&1&0&0&1&0\\
1&0&1&0&1&0&0\\
0&1&1&1&0&0&0\\
\end{array}
\right]= G
,
\end{displaymath}
\noindent
which is the generator matrix of the Hamming code $\mathcal{H}(7, 4, 3)$. Note that $7-\log_2(1+7)=4=\dim \mathbb{V}$. Figure 1 shows the implementation of the binary Hamming Transform $7 \times 7$.
%
%
\begin{figure}[ht]
\centering
\includegraphics[scale=0.5]{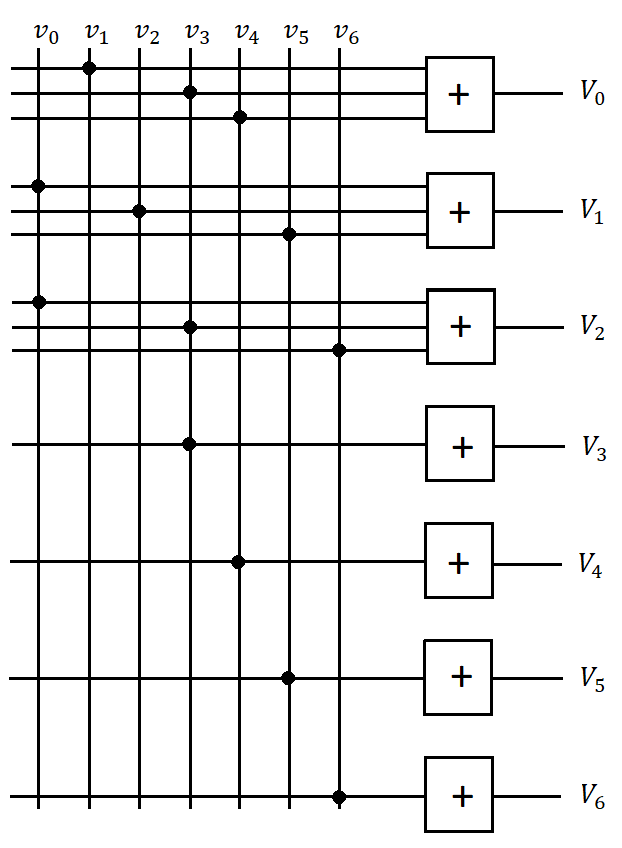}
\caption{Block diagram for the binary Hamming Transform of length 7.}
\label{fig:HammingTransform}
\end{figure}
\begin{definition}
\label{def2}
The Hamming Number-Theoretic Transform, with length $N=(p^m-1)/(p-1)$, of the sequence $\textbf{v}=(v_0,v_1,\cdots,v_{N-1})^T$, $ v_i \in GF(p)$, is the sequence $\textbf{V}=(V_0,V_1,\cdots,V_{N-1})^T,V_k \in GF(p)$, given by
\begin{equation}
\textbf{V}=T_{HAM}^{(\lambda)} \cdot \textbf{v},
\end{equation}
\noindent where $T_{HAM}^{(\lambda)}$ is the Hamming transformation matrix over $GF(p)$, parametrized by eigenvalue $\lambda$.
\end{definition}
The only natural property is that this transformation is linear, since the distributive property of multiplication relative to addition holds by matrices. Furthermore, all the codewords of the classic Hamming code are invariant under the HamNT.
\begin{ex}
(nonbinary Hamming transform). Consider the ternary Hamming code $\mathcal{H}(13,10,3)$ whose parity-check matrix is 
\begin{equation*}
\arraycolsep=2pt
H=\left[ 
\begin{array}{lllllllllllll}
0 & 0 & 0 & 0 & 1 & 1 & 1 & 1 & 1 & 1 & 1 & 1 & 1 \\ 
0 & 1 & 1 & 1 & 0 & 0 & 0 & 1 & 1 & 1 & 2 & 2 & 2 \\ 
1 & 0 & 1 & 2 & 0 & 1 & 2 & 0 & 1 & 2 & 0 & 1 & 2 \\
\end{array}\right].
\end{equation*}
The transform matrix of the ternary Hamming transform, with eigenvalue $\lambda=1$, is given by
\begin{equation*}
\arraycolsep=2pt
\renewcommand{\arraystretch}{0.9}
T_{HAM}^{(1)} = \left[ \begin{array}{lllllllllllll}
1 & 0 & 0 & 0 & 1 & 1 & 1 & 1 & 1 & 1 & 1 & 1 & 1 \\ 
0 & 2 & 1 & 1 & 0 & 0 & 0 & 1 & 1 & 1 & 2 & 2 & 2 \\ 
1 & 0 & 2 & 2 & 0 & 1 & 2 & 0 & 1 & 2 & 0 & 1 & 2 \\ 
0&0&0&1&0&0&0&0&0&0&0&0&0\\
0&0&0&0&1&0&0&0&0&0&0&0&0\\
0&0&0&0&0&1&0&0&0&0&0&0&0\\
0&0&0&0&0&0&1&0&0&0&0&0&0\\
0&0&0&0&0&0&0&1&0&0&0&0&0\\
0&0&0&0&0&0&0&0&1&0&0&0&0\\
0&0&0&0&0&0&0&0&0&1&0&0&0\\
0&0&0&0&0&0&0&0&0&0&1&0&0\\
0&0&0&0&0&0&0&0&0&0&0&1&0\\
0&0&0&0&0&0&0&0&0&0&0&0&1\\
\end{array}\right].
\end{equation*}
\end{ex}
\vspace{6pt}

This kind of construction makes it harder to find the properties of such a transform. In what follows we adopt a polynomial approach.

\subsection{The Cyclic Hamming Transform}

Another way to represent the Hamming transform over $GF(p)$ introduced in the previous subsection is considering the parity-check matrix $H$ as expressed in the form

\begin{displaymath}
H=\left[\begin{array}{c}
h(x)\\ 
xh(x)\\
x^2h(x)\\
\vdots\\
x^{n-k-1}h(x)
\end{array}\right],
\end{displaymath}

\noindent where $h(x)$ is the parity polynomial of the cyclic Hamming code over $GF(p)$ \cite{Moon2005}. The $k$ rows necessary to construct the matrix $H_e$ are generated by cyclic shifting the parity polynomial, $h(x)$. 

\begin{ex}
Consider the binary Hamming cyclic code $\mathcal{H}(7,4,3)$ with parity polynomial given by $h(x)=x^4+x^2+x+1$. The parity-check matrix $H$ is given by

\begin{displaymath}
\arraycolsep=2pt
\renewcommand{\arraystretch}{1}
H=
\left[ \begin{array}{lllllll}
1&0&1&1&1&0&0\\
0&1&0&1&1&1&0\\
0&0&1&0&1&1&1\\
\end{array}
\right].
\end{displaymath}

\noindent The transformation matrix, in this case, is

\begin{displaymath}
\arraycolsep=2pt
\renewcommand{\arraystretch}{1}
T_{HAM}^{(\lambda)}=
\left[ \begin{array}{lllllll}
1&0&1&1&1&0&0\\
0&1&0&1&1&1&0\\
0&0&1&0&1&1&1\\
1&0&0&1&0&1&1\\
1&1&0&0&1&0&1\\
1&1&1&0&0&1&0\\
0&1&1&1&0&0&1\\
\end{array}
\right]+\lambda \mathbb{I}_7.
\end{displaymath}
\noindent Taking $\lambda=1$ results in the cyclic form of the binary Hamming transformation matrix (the inverse matrix is also circulant)
\begin{displaymath}
\arraycolsep=2pt
\renewcommand{\arraystretch}{1}
\tilde{T}_{HAM}^{(1)} = 
\left[ \begin{array}{lllllll}
0&0&1&1&1&0&0\\
0&0&0&1&1&1&0\\
0&0&0&0&1&1&1\\
1&0&0&0&0&1&1\\
1&1&0&0&0&0&1\\
1&1&1&0&0&0&0\\
0&1&1&1&0&0&0\\
\end{array}
\right].
\end{displaymath}
\hfill{$\blacksquare$}
\end{ex}
The transform matrices $T_{HAM}^{(1)}$ and $\tilde{T}_{HAM}^{(1)}$, ilustrated in Fig.2, are different representations for the Hamming transform over $GF(2)$. Note that, by considering the rows of $\lambda \mathbb{I}_N$ as cyclic shifts of the vector $[\lambda 0 0 \cdots 0]$, in polynomial notation, we can write

\begin{displaymath}
\renewcommand{\arraystretch}{1}
\lambda \mathbb{I}_N=\left[\begin{array}{c}
\lambda(x)\\ 
x\lambda(x)\\
x^2\lambda(x)\\
\vdots\\
x^{N-1}\lambda
\end{array}\right].
\end{displaymath}

\begin{figure}[ht]
\centering
\includegraphics[scale=0.36]{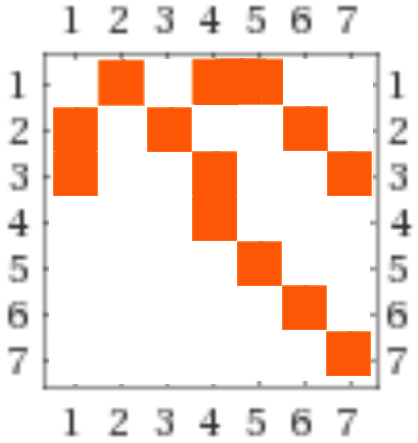}
\includegraphics[scale=0.40]{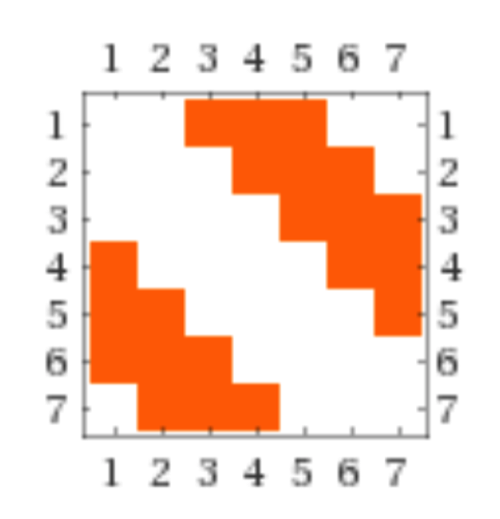}
\caption{Illustration of Hamming Transforms: $T_{HAM}^{(1)}, \tilde{T}_{HAM}^{(1)}$.}
\label{fig:HammingTransform2}
\end{figure}

%
\begin{definition}
\label{def3}
The Cyclic Hamming Number-Theoretic Transform  (CHamNT), with length $N=(p^{m}-1)/(p-1)$,  of the sequence $\bf{v}= (v_0,v_1,\cdots,v_{N-1})^T$, $v_i \in GF(p)$, is the sequence $\bf{V}= (V_0,V_1,\cdots,V_{N-1})^T$, $V_k \in GF(p)$, such that $\bf{V}=\tilde{T}_{HAM}^{(\lambda)}\cdot \bf{v}$, where 
\begin{center}
\begin{equation*}
T_{HAM}^{(\lambda)}(x) =
\left[\begin{array}{c}
[h(x)+\lambda(x)]\\ 
x[h(x)+\lambda(x)]\\
x^2[h(x)+\lambda(x)]\\
\vdots\\
x^{N-1}[h(x)+\lambda(x)]
\end{array}\right].
\end{equation*}
\end{center}
\hfill{$\blacksquare$}
\end{definition}
\vspace{6pt}
\noindent Note that this definition paves the way for an algebraic treatment of the properties of the CHamNT.\vspace{6pt}

\begin{prop}
\label{prop6_1}
The matrix $T_{HAM}^{(1)}$ is a circulant matrix.
\end{prop}
\vspace{6pt}
The procedure to generate $H_e$ by means of cyclic shifts of the parity polynomial, $h(x)$, results in a circulant matrix. This aspect is not modified by adding $\lambda \mathbb{I}_N$ to this matrix.

\subsection{Some of Properties of the Cyclic Hamming Number-Theoretic Transform}

\begin{enumerate}[label=\roman*)]
\item \textbf{Linearity}
\item \textbf{Time shift}\\ 
Consider the sequence $\widehat{\bf{v}}=(\widehat{v}_0,\cdots,\widehat{v}_{N-1})$ where $\widehat{v}_i=v_{i-m}.$ Then, $\widehat{\bf{v}} \leftrightarrow \widehat{\bf{V}}$, where
\begin{equation}
\widehat{\bf{V}}=x^{m}\bf{V}.
\end{equation}
\begin{proof}
Note  that we are considering, without loss of generality, a cyclic shift of $m$ positions to the right. Therefore, we have $\widehat{\bf{v}}=x^{m}\bf{v}$ ~\textit{(mod $x^{N}-1$)}. Then,
\begin{eqnarray*}
\renewcommand{\arraystretch}{0.7}
\widehat{\bf{V}}=
\left[ \begin{array}{@{}c@{}}
[h(x)+\lambda(x)]\\ 
x[h(x)+\lambda(x)]\\
x^2[h(x)+\lambda(x)]\\
\vdots\\
x^{N-1}[h(x)+\lambda(x)]
\end{array}
\right]
\left[ \begin{array}{c}
x^{m}v_0\\
x^{m}v_1\\
x^{m}v_2\\
\vdots\\
x^{m}v_{N-1}\\
\end{array}
\right]
\end{eqnarray*}
\begin{eqnarray*}
\renewcommand{\arraystretch}{0.7}
&=&x^{m} \left[\begin{array}{c}
[h(x)+\lambda(x)]\\ 
x[h(x)+\lambda(x)]\\
x^2[h(x)+\lambda(x)]\\
\vdots\\
x^{N-1}[h(x)+\lambda(x)]
\end{array}
\right] 
\left[\begin{array}{c}
v_0\\
v_1\\
\vdots\\
v_{N-1}
\end{array}
\right]\\
&=&x^{m}\bf{V}.
\end{eqnarray*}
\end{proof}
\item \textbf{Frequency shift}\\ 
Consider the sequence $\widehat{\bf{V}}=(\widehat{V}_0,\cdots,\widehat{V}_{N-1})$ where $\widehat{V}_k=V_{k-l}. $ Then, $\widehat{\bf{v}}=x^{l}\bf{v}.$
The demonstration is similar to the previous one.
\item \textbf{Constant sequence transform}\\
The transform of a sequence $\bf{v}= (\it{r,r,\cdots,r})$ is the sequence $V_k = r\cdot weight(h(x))$, $\forall k$.

This property means that the Hamming transform of a constant vector is also a constant vector. This contrasts sharply with the discrete Fourier transform, where there is a trade-off between the two domains and compressing into one domain implies expanding into the other.

\item \textbf{Impulse sequence transform}\\
The transform of the sequence $\delta = (1,0,\cdots,0)$, corresponds to the first column of the matrix $T_{H}^{(\lambda)}$, i.e., the coefficients of $x[h(x)+\lambda(x)]$.
\end{enumerate}

\section{The Golay Number-Theoretic Transform}
There exist two Golay codes, the binary $\mathcal{G}(23,12,7)$ $(h(x)=x^{12}+x^{11}+x^{10}+x^9+x^8+x^5+x^2+1)$ and the ternary $\mathcal{G}(11,6,5)$ $(h(x)=x^6+2x^5+2x^4+2x^3+x^2+1)$.

Like in the previous section, to define the Golay number theoretic transform, we adopt a polynomial approach.

\subsection{The Binary Golay Transform}
The construction of the transformation matrix of the Cyclic Golay Transform follows the same steps used in the construction of the Cyclic Hamming Transform. Note that $2^{12} \left [ 1+23+253+1771 \right ] = 2^{23}$ and that $23-\log_2(1+23+253+1771)=12=\dim \mathbb{V}$.

The Golay binary $\mathcal{G}(23,12,7)$ code has parity polynomial $h(x)=x^{12}+x^{11}+x^{10}+x^9+x^8+x^5+x^2+1$, so that\vspace{6pt}
\begin{equation*}
\arraycolsep=2pt
T_{CGolNT}^{(\lambda)}(x) =
\begin{bmatrix}
\footnotesize
\setlength\tabcolsep{1pt}
\begin{tabular}{lllllllllllllllllllllll}
0 & 1 & 1 & 1 & 1 & 0 & 0 & 1 & 0 & 0 & 1 & 0 & 1 & 0 & 0 & 0 & 0 & 0 & 0 & 0 & 0 & 0 & 0  \\ 
0 & 0 & 1 & 1 & 1 & 1 & 0 & 0 & 1 & 0 & 0 & 1 & 0 & 1 & 0 & 0 & 0 & 0 & 0 & 0 & 0 & 0 & 0  \\ 
0 & 0 & 0 & 1 & 1 & 1 & 1 & 0 & 0 & 1 & 0 & 0 & 1 & 0 & 1 & 0 & 0 & 0 & 0 & 0 & 0 & 0 & 0  \\ 
0 & 0 & 0 & 0 & 1 & 1 & 1 & 1 & 0 & 0 & 1 & 0 & 0 & 1 & 0 & 1 & 0 & 0 & 0 & 0 & 0 & 0 & 0  \\ 
0 & 0 & 0 & 0 & 0 & 1 & 1 & 1 & 1 & 0 & 0 & 1 & 0 & 0 & 1 & 0 & 1 & 0 & 0 & 0 & 0 & 0 & 0  \\ 
0 & 0 & 0 & 0 & 0 & 0 & 1 & 1 & 1 & 1 & 0 & 0 & 1 & 0 & 0 & 1 & 0 & 1 & 0 & 0 & 0 & 0 & 0  \\ 
0 & 0 & 0 & 0 & 0 & 0 & 0 & 1 & 1 & 1 & 1 & 0 & 0 & 1 & 0 & 0 & 1 & 0 & 1 & 0 & 0 & 0 & 0  \\ 
0 & 0 & 0 & 0 & 0 & 0 & 0 & 0 & 1 & 1 & 1 & 1 & 0 & 0 & 1 & 0 & 0 & 1 & 0 & 1 & 0 & 0 & 0  \\ 
0 & 0 & 0 & 0 & 0 & 0 & 0 & 0 & 0 & 1 & 1 & 1 & 1 & 0 & 0 & 1 & 0 & 0 & 1 & 0 & 1 & 0 & 0  \\ 
0 & 0 & 0 & 0 & 0 & 0 & 0 & 0 & 0 & 0 & 1 & 1 & 1 & 1 & 0 & 0 & 1 & 0 & 0 & 1 & 0 & 1 & 0  \\ 
0 & 0 & 0 & 0 & 0 & 0 & 0 & 0 & 0 & 0 & 0 & 1 & 1 & 1 & 1 & 0 & 0 & 1 & 0 & 0 & 1 & 0 & 1  \\ 
1 & 0 & 0 & 0 & 0 & 0 & 0 & 0 & 0 & 0 & 0 & 0 & 1 & 1 & 1 & 1 & 0 & 0 & 1 & 0 & 0 & 1 & 0  \\ 
0 & 1 & 0 & 0 & 0 & 0 & 0 & 0 & 0 & 0 & 0 & 0 & 0 & 1 & 1 & 1 & 1 & 0 & 0 & 1 & 0 & 0 & 1  \\ 
1 & 0 & 1 & 0 & 0 & 0 & 0 & 0 & 0 & 0 & 0 & 0 & 0 & 0 & 1 & 1 & 1 & 1 & 0 & 0 & 1 & 0 & 0  \\ 
0 & 1 & 0 & 1 & 0 & 0 & 0 & 0 & 0 & 0 & 0 & 0 & 0 & 0 & 0 & 1 & 1 & 1 & 1 & 0 & 0 & 1 & 0  \\ 
0 & 0 & 1 & 0 & 1 & 0 & 0 & 0 & 0 & 0 & 0 & 0 & 0 & 0 & 0 & 0 & 1 & 1 & 1 & 1 & 0 & 0 & 1  \\ 
1 & 0 & 0 & 1 & 0 & 1 & 0 & 0 & 0 & 0 & 0 & 0 & 0 & 0 & 0 & 0 & 0 & 1 & 1 & 1 & 1 & 0 & 0  \\ 
0 & 1 & 0 & 0 & 1 & 0 & 1 & 0 & 0 & 0 & 0 & 0 & 0 & 0 & 0 & 0 & 0 & 0 & 1 & 1 & 1 & 1 & 0  \\ 
0 & 0 & 1 & 0 & 0 & 1 & 0 & 1 & 0 & 0 & 0 & 0 & 0 & 0 & 0 & 0 & 0 & 0 & 0 & 1 & 1 & 1 & 1  \\ 
1 & 0 & 0 & 1 & 0 & 0 & 1 & 0 & 1 & 0 & 0 & 0 & 0 & 0 & 0 & 0 & 0 & 0 & 0 & 0 & 1 & 1 & 1  \\ 
1 & 1 & 0 & 0 & 1 & 0 & 0 & 1 & 0 & 1 & 0 & 0 & 0 & 0 & 0 & 0 & 0 & 0 & 0 & 0 & 0 & 1 & 1  \\ 
1 & 1 & 1 & 0 & 0 & 1 & 0 & 0 & 1 & 0 & 1 & 0 & 0 & 0 & 0 & 0 & 0 & 0 & 0 & 0 & 0 & 0 & 1  \\ 
1 & 1 & 1 & 1 & 0 & 0 & 1 & 0 & 0 & 1 & 0 & 1 & 0 & 0 & 0 & 0 & 0 & 0 & 0 & 0 & 0 & 0 & 0  \\ 
\end{tabular}
\end{bmatrix},
\end{equation*}

\noindent the determinant of which is equals to one.
\subsection{The Ternary Golay Transform}
The Golay ternary parity polynomial is $h(x)=x^6+2x^5+2x^4+2x^3+x^2+1$. Considering $\lambda = 1$ as eigenvalue, we have the following circulant form for the Golay transform matrix:
\begin{equation*}
\scriptsize
\arraycolsep=2pt
T_{CGolNT}^{(1)} =
\left[\begin{array}{lllllllllll}
2&2&2&2&1&0&1&0&0&0&0\\
0&2&2&2&2&1&0&1&0&0&0\\
0&0&2&2&2&2&1&0&1&0&0\\
0&0&0&2&2&2&2&1&0&1&0\\
0&0&0&0&2&2&2&2&1&0&1\\
1&0&0&0&0&2&2&2&2&1&0\\
0&1&0&0&0&0&2&2&2&2&1\\
1&0&1&0&0&0&0&2&2&2&2\\
2&1&0&1&0&0&0&0&2&2&2\\
2&2&1&0&1&0&0&0&0&2&2\\
2&2&2&1&0&1&0&0&0&0&2\\
\end{array}\right].
\end{equation*}
The ternary eigenvector space of the Golay transform consists of $3^6=729$ vectors \cite{Pless 1968}. According to Sagemath\textsuperscript{\textregistered}, this transformation matrix has multiplicative order 242, characteristic polynomial $(2+x)^6(1+x+x^2+x^3+2x^4+x^5)$ and $\lambda=1$ is an eigenvalue with algebraic multiplicity 6.

In the next example, we consider the construction of the ternary Golay transform starting from the systematic form of the Golay  code parity-check matrix.
\begin{ex}
Consider the ternary Golay code $\mathcal{G}(11,6,5)$ whose parity-check matrix is
\begin{equation*}
\scriptsize
\arraycolsep=2pt
H=
\left[\begin{array}{lllllllllll}
1&1&1&2&2&0&1&0&0&0&0\\
1&1&2&1&0&2&0&1&0&0&0\\
1&2&1&0&1&2&0&0&1&0&0\\
1&2&0&1&2&1&0&0&0&1&0\\
1&0&2&2&1&1&0&0&0&0&1\\
\end{array}\right].
\end{equation*}
We inflate this matrix to a square matrix by adding null rows. 
\noindent To this matrix we add $\lambda \mathbb{I}_{11}$. Assuming $\lambda=1$, it leads to \vspace{6pt}
\begin{equation*}
\scriptsize
\arraycolsep=2pt
T_{GolNT}^{(1)} = 
\left[\begin{array}{lllllllllll}
2&1&1&2&2&0&1&0&0&0&0\\
1&2&2&1&0&2&0&1&0&0&0\\
1&2&2&0&1&2&0&0&1&0&0\\
1&2&0&2&2&1&0&0&0&1&0\\
1&0&2&2&2&1&0&0&0&0&1\\
0&0&0&0&0&1&0&0&0&0&0\\
0&0&0&0&0&0&1&0&0&0&0\\
0&0&0&0&0&0&0&1&0&0&0\\
0&0&0&0&0&0&0&0&1&0&0\\
0&0&0&0&0&0&0&0&0&1&0\\
0&0&0&0&0&0&0&0&0&0&1\\
\end{array}\right].
\end{equation*}
\noindent This transform has the characteristic polynomial $p(x) =1 + 2x^3 + 2x^4 + x^5 + 2x^6 + 2x^7+ x^8 + x^9 + 2x^{10}+x^{11}$, eigenvalue $\lambda=1$ and determinant equals to 2. In this transform, the eigenvector subspace associated with $\lambda=1$ has dimension $\dim \mathbb{V}=11-\log_3 (1+22+220)=6$. Note that $T_{CGolNT}^{(1)}$ and $T_{GolNT}^{(1)}$ are different representations for the Golay transform of length 11 over $GF(3)$, the first one being in the form of a circulating matrix.
\end{ex}

\subsection{A few properties of the cyclic Golay Number-Theoretic Transform}
\begin{enumerate}[label=\roman*)]
\item \textbf{Linearity}
\item \textbf{Time shift}\\
Consider the sequence $\widehat{\bf{v}}=(\widehat{v}_0,\cdots,\widehat{v}_{N-1})$ where $\widehat{v}_i=v_{i-m}.$ Then, $\widehat{\bf{v}} \leftrightarrow \widehat{\bf{V}}$, where
\begin{displaymath}
\widehat{\bf{V}}=x^{m}\bf{V}.
\end{displaymath}
\item \textbf{Frequency shift}\\ 
Consider the sequence $\widehat{\bf{V}}=(\widehat{V}_0,\cdots, \widehat{V}_{N-1})$ where $\widehat{V}_k=V_{k-l}. $ Then, $\widehat{\bf{v}}=x^{l}\bf{v}.$
\item \textbf{Constant sequence transform}\\
The transform of the sequence $\bf{v}= (\it{r,r,\cdots,r})$ is the sequence $V_k = r\cdot weight(h(x))$, $\forall k$.

\item \textbf{Impulse sequence transform}\\
The transform of the sequence $\delta = (1,0,\cdots,0)$, corresponds to the first column of the matrix $T^{(\lambda)}_{CGolNT}$, i.e., the coefficients of the polynomial $x[h(x)+\lambda(x)]$.
\end{enumerate}
These properties hold for both the binary and ternary Golay transform, simply using the corresponding parity polynomial.

\section{Golay Extended Transform}
The extended versions of the Hamming or Golay codes, which are $selfdual$ codes \cite{Conway-Sloane 2013}, can be used to construct new transforms. For the ternary Golay codes, the extended Golay code has parameters $\mathcal{G}(N=12, k=6, d=6)$ over $GF(3)$. One possible approach over this field is to take into account that $2 \equiv -1$ and the transform matrix can be viewed only as computing additions (multiplication free transform). It should be interesting to compare it with Hadamard transforms \cite{Horadam 2012}. The extended Golay parity-check matrix is given by
\begin{equation*}
\footnotesize
\arraycolsep=2pt
H=
\left[\begin{array}{rrrrrrrrrrrr}
0&-1&-1&-1&-1&-1&1&0&0&0&0&0\\
-1&0&-1&1&1&-1&0&1&0&0&0&0\\
-1&-1&0&-1&1&1&0&0&1&0&0&0\\
-1&1&-1&0&-1&1&0&0&0&1&0&0\\
-1&1&1&-1&0&1&0&0&0&0&1&0\\
-1&-1&1&1&-1&0&0&0&0&0&0&1\\
\end{array}\right].
\end{equation*}

\noindent Inflating this matrix to a square matrix by adding the following (arbitrary) linear combinations: $l_1+l_2,\ l_1+l_3,\ l_1+l_4,\ l_1+l_5,\ l_1+l_6$ e $l_2+l_3,$ then adding $\lambda \mathbb{I}_{12}$, and assuming $\lambda=1$, results in the $12 \times 12$ transform matrix  
%
%
%
\begin{equation*}
\footnotesize
\arraycolsep=2pt
T_{EG}^{(1)}  =
\left[\begin{array}{rrrrrrrrrrrr}
1&-1&-1&-1&-1&-1&1&0&0&0&0&0\\
-1&1&-1&1&1&-1&0&1&0&0&0&0\\
-1&-1&1&-1&1&1&0&0&1&0&0&0\\
-1&1&-1&1&-1&1&0&0&0&1&0&0\\
-1&1&1&-1&1&1&0&0&0&0&1&0\\
-1&-1&1&1&-1&1&0&0&0&0&0&1\\
-1&-1&1&0&0&1&-1&1&0&0&0&0\\
-1&1&-1&1&0&0&1&1&1&0&0&0\\
-1&0&1&-1&1&0&1&0&1&1&0&0\\
-1&0&0&1&-1&0&1&0&0&1&1&0\\
-1&1&0&0&1&-1&1&0&0&0&1&1\\
1&-1&-1&0&-1&0&0&1&1&0&0&1\\
\end{array}\right],
\end{equation*}
i.e., the matrix of the ternary extended Golay transform ($\text{det }T_{EG}^{(1)}=2$). This new transform requires only additions/subtractions in order to be computed. Their symmetries properties inherited from the selfdual structure of the space of eigenvectors can be attractive. The inverse transform matrix $T_{EG}^{-1}$ is
\begin{equation*}
\footnotesize
\arraycolsep=2pt
\left[\begin{array}{rrrrrrrrrrrr}
1&0&-1&1&1&1&0&-1&1&1&1&1\\
-1&1&0&1&-1&0&0&-1&1&1&0&0\\
0&1&-1&0&1&-1&-1&-1&1&-1&0&1\\
0&1&0&0&0&-1&1&-1&-1&1&-1&-1\\
1&1&0&-1&1&1&0&-1&1&0&-1&0\\
-1&-1&-1&0&0&0&1&1&1&-1&1&-1\\
-1&0&-1&-1&0&1&1&1&-1&-1&1&1\\
0&1&0&1&-1&0&-1&-1&-1&0&1&-1\\
0&1&-1&0&1&0&1&1&1&-1&0&0\\
1&1&-1&0&1&-1&0&1&-1&-1&0&1\\
-1&-1&1&1&1&0&1&0&-1&0&0&0\\
-1&1&1&1&0&-1&-1&1&-1&0&0&-1\\
\end{array}\right],
\end{equation*}
Finally, it is worth mentioning that since the eigenvalue is unity for the direct transform, all eigenvectors of the inverse transform are the same for the direct transform.
\section{Final Remarks}
The introduction of the Hamming (HamNT) and Golay (GolNT) number-theoretic transforms represents an important application of the theory introduced in \cite{Campello de Souza 2009}, \cite{Campello de Souza 2011} and outlines the existence of some kind of $isomorphism$ between linear block codes and finite field transforms. All the codewords of the classic Hamming code are invariant under HamNT (the same happens for the Golay code and the GolNT). A probable link between the Golay transform and Mathieu's simple group deserves to be investigated \cite{Conway-Sloane 2013}, \cite{Thompson 1983}. These transforms can be used in Image Processing \cite{Boussakta-Holt 1999}. Even if it is an introductory work of presenting new ideas, without investigating potential applications, the strength of geometric configurations linked to perfect codes \cite{Berlekamp 1973}, \cite{Kimizura-Sasaki 2008} seems to give an inkle of the potential of these new transforms. This also opens up interesting perspectives in designing fast algorithms for the computation of these transformations \cite{Amira 2001}, \cite{Meher 2008}.
\section*{Acknowledgements}
The authors would like to thank the support received from their institutions.

\appendix
\section{}
\noindent
A general formulation of a discrete transform from systematic block codes: In the general case, starting from a block code with parity-check matrix 
\begin{equation*}
 H=
\bigg[\begin{array}{c|c}
-\mathbb{P}_{(N-k) \times k}^T & \mathbb{I}_{(N-k) \times (N-k)} 
\end{array}\bigg]   
\end{equation*}

\noindent we arrive at a square transform matrix, $N \times N$, invertible, given by\\
\begin{equation*}
T_{\text{block code}} = 
\bigg[\begin{array}{c|c}
\lambda \mathbb{I}_{(N-k) \times k}-\mathbb{P}_{(N-k) \times k}^T & \mathbb{I}_{(N-k) \times (N-k)}\\ \hline
\mathbb{O}_{k \times {(N-k)} } & 
\lambda \mathbb{I}_{k \times k}
\end{array}\bigg].
\end{equation*}
\end{document}